\documentstyle[12pt]{article}
\begin{document}

\begin{flushright}{UT-02-07\\IU-MSTP/46}
\end{flushright}
\vskip 0.5 truecm

\begin{center}
{\Large{\bf  Ginsparg-Wilson operators and a no-go theorem
 }}
\end{center}
\vskip .5 truecm
\centerline{\bf Kazuo Fujikawa,  Masato Ishibashi}
\vskip .4 truecm
\centerline {\it Department of Physics,University of Tokyo}
\centerline {\it Bunkyo-ku,Tokyo 113,Japan}
\vskip 0.5 truecm
\centerline {\it and}
\vskip 0.5 truecm
\centerline{\bf Hiroshi Suzuki}
\vskip .4 truecm
\centerline {\it Department of Mathematical Sciences, Ibaraki 
University}
\centerline {\it Mito, 310-8512, Japan}

\begin{abstract}
If one uses a general class of 
Ginsparg-Wilson operators, it is known that CP symmetry is 
spoiled in chiral gauge theory for a finite lattice spacing and 
the Majorana  
fermion is not defined in the presence of chiral symmetric 
Yukawa couplings.
We summarize these properties in the form of  a theorem for the 
general Ginsparg-Wilson relation.
\end{abstract}


The Ginsparg-Wilson relation\cite{ginsparg} provides a 
convenient framework for the analyses of chiral symmetry in 
lattice theory\cite{niedermayer}. 
It has been recently pointed out by Hasenfratz\cite{hasenfratz} 
that the overlap operator\cite{neuberger} has a conflict with CP 
symmetry in chiral gauge theory for any finite lattice spacing 
$a$. We pointed out that the lattice chiral symmetry of the 
Ginsparg-Wilson operator 
has a certain conflict with the definition of the Majorana 
fermion in the presence of Yukawa couplings\cite{fuji-ishi}. 
(The breaking of CP symmetry in a different context was also
mentioned in\cite{fuji-ishi}).
These analyses are either based on the simple form of the 
Ginsparg-Wilson relation and its explicit 
solution\cite{hasenfratz}, or on the generalized forms of 
Ginsparg-Wilson relation but still on their explicit 
solutions\cite{fuji-ishi}. 
It may be useful to formulate these properties in a more 
abstract and general setting to understand the general features 
of these complications.
In this paper we provide such an analysis. 

We deal with a hermitian lattice operator defined by
\begin{equation}
H=a\gamma_{5}D=H^{\dagger}=aD^{\dagger}\gamma_{5}
\end{equation}
where $D$ stands for the lattice Dirac operator and $a$ is the 
lattice spacing. 
We analyze the 
general Dirac operator defined by the algebraic 
relation
\begin{equation}
\gamma_{5}H+H\gamma_{5}=2H^{2}f(H^{2})
\end{equation}
where $f(H^{2})$ is assumed to be a regular function of $H^{2}$ 
and $f(H^{2})^{\dagger}=f(H^{2})$: To be definite, we assume 
that $f(x)$ is monotonous and non-decreasing for $x\geq 0$, 
and  $f(H^{2})=1$ corresponds to the conventional Ginsparg-Wilson
relation\cite{niedermayer}.
We also assume that the operator $H$ is local in the sense that 
it is analytic in the entire Brillouin zone.
One can confirm the relation
\begin{equation}
\gamma_{5}H^{2}=(\gamma_{5}H+H\gamma_{5})H-
H(\gamma_{5}H+H\gamma_{5})+ H^{2}\gamma_{5}=H^{2}\gamma_{5}
\end{equation}
which implies $H^{2}=\gamma_{5}H^{2}\gamma_{5}$ and thus 
$DH^{2}=H^{2}D$. 
The above defining relation (2) is written in a variety of ways
such as  
\begin{eqnarray}
&&\Gamma_{5}H+H\Gamma_{5}=0,\ \ \ 
\gamma_{5}\Gamma_{5}\gamma_{5}D + D\Gamma_{5}=0,\nonumber\\
&&\gamma_{5}H+H\hat{\gamma}_{5}=0,\ \ \ 
\gamma_{5}D + D\hat{\gamma}_{5}=0
\end{eqnarray}
 and $\hat{\gamma}^{2}_{5}=1$ where
\begin{eqnarray}
&&\Gamma_{5}=\gamma_{5}-Hf(H^{2}),\nonumber\\
&&\hat{\gamma}_{5}=\gamma_{5}-2Hf(H^{2}).
\end{eqnarray}
We also note the relation
\begin{eqnarray}
\gamma_{5}\Gamma_{5}\hat{\gamma}_{5}&=&\gamma_{5}2\Gamma^{2}_{5}
-\gamma_{5}\Gamma_{5}\gamma_{5}=\gamma_{5}(\gamma_{5}\Gamma_{5}
+\Gamma_{5}\gamma_{5})-\gamma_{5}\Gamma_{5}\gamma_{5}\nonumber\\
&=&\gamma_{5}(\gamma_{5}\Gamma_{5}).
\end{eqnarray}

We now examine the action defined by
\begin{equation}
S=\int d^{4}x\bar{\psi}D\psi\equiv\sum_{x,y}\bar{\psi}(x)D(x,y)
\psi(y)
\end{equation}
which is invariant under the lattice chiral transformation
\begin{equation}
\delta\psi=i\epsilon\hat{\gamma}_{5}\psi, \ \ 
\delta\bar{\psi}=\bar{\psi}i\epsilon\gamma_{5}.
\end{equation}
If one considers the field re-definition
\begin{equation}
\psi^{\prime}=\gamma_{5}\Gamma_{5}\psi,\ \ \ 
\bar{\psi}^{\prime}=\bar{\psi}
\end{equation}
the above action is written as 
\begin{equation}
S=\int d^{4}x\bar{\psi}^{\prime}D\frac{1}{\gamma_{5}\Gamma_{5}}
\psi^{\prime}
\end{equation}
which is invariant under the naive chiral transformation (by 
using (6))
\begin{eqnarray}
\delta\psi^{\prime}&=&\gamma_{5}\Gamma_{5}\delta\psi
=\gamma_{5}\Gamma_{5}i\epsilon\hat{\gamma}_{5}\psi
=i\epsilon\gamma_{5}\psi^{\prime},\nonumber\\
\delta\bar{\psi}^{\prime}&=&\bar{\psi}^{\prime}i\epsilon
\gamma_{5}.
\end{eqnarray}
This chiral symmetry implies the relation
\begin{equation}
\{\gamma_{5},D\frac{1}{\gamma_{5}\Gamma_{5}} \}=0.
\end{equation}
We here recall the conventional no-go theorem in the form of 
Nielsen and Ninomiya\cite{nielsen}, which states in view of (10)
and (12) that\\
(i) If the operator $D$ is local and if
$1/(\gamma_{5}\Gamma_{5})$ is analytic in the entire 
Brillouin zone, the operator $D$ contains the species 
doubling. The simplest choice $f(H^{2})=0$ and thus 
$\Gamma_{5}=\gamma_{5}$ is included in this  case.\\
(ii)If the operator $D$ is local and free of species doubling,
then the operator $\gamma_{5}\Gamma_{5}$ is also local by 
its construction. But the operator 
$1/(\gamma_{5}\Gamma_{5})$ cannot be analytic in 
the entire Brillouin zone, which in turn suggests that 
\begin{equation}
\Gamma^{2}_{5}=1-H^{2}f^{2}(H^{2})=0
\end{equation}
has solutions inside the Brillouin zone. These properties are 
proved for vanishing gauge field.

Since we are interested in the local and doubler-free 
operator $D$, we can summarize the analysis so far in the form 
of a  theorem.\\ {\bf Theorem}:
For any lattice operator $D$ defined by the algebraic relation 
(2), which is local (i.e., analytic in the entire Brillouin zone)
and free of species doubling, the operator 
$1/(\gamma_{5}\Gamma_{5})$ is singular inside the Brillouin 
zone and $\Gamma^{2}_{5}=1-H^{2}f^{2}(H^{2})$ has at least one 
zero inside the Brillouin zone. 
 
These properties are known\cite{fuji-ishi} for the 
specific case $f(H^{2})=H^{2k}$ with a 
non-negative integer $k$ by using explicit 
solutions\cite{fujikawa}. 
Here we learn that these properties are of more general 
validity and intrinsically related to the basic notions of 
locality and species doubling. 
\\

In the following we discuss the implications of the above 
theorem.  
We first summarize the representation of the algebra (2). See,
for example, ref.\cite{fujikawa}.
Let us consider the eigenvalue problem
\begin{equation}
H\varphi_n(x)=\lambda_n\varphi_n(x),\quad(\varphi_n,\varphi_m)
=\delta_{nm}.
\end{equation}
We first note
$H\Gamma_5\varphi_n(x)=-\Gamma_5H\varphi_n(x)
=-\lambda_n\Gamma_5\varphi_n(x)$,
and
\begin{equation}
(\Gamma_5\varphi_n,\Gamma_5\varphi_m)
=[1-\lambda_n^2f^2(\lambda_n^2)]\delta_{nm}.
\end{equation}
These relations show that eigenfunctions with $\lambda_n\neq0$ 
and $\lambda_nf(\lambda_n^2)\neq\pm1$ come in pairs as 
$\lambda_n$ and $-\lambda_n$ (when$\lambda_n=0$,
$\varphi_0(x)$ and $\Gamma_5\varphi_0(x)$ are not necessarily 
linearly independent).

We can thus classify eigenfunctions as follows:\\
\noindent
(i) $\lambda_n=0$ ($H\varphi_0(x)=0$). For this
one may impose the chirality on $\varphi_0(x)$ as
\begin{equation}
\gamma_5\varphi_0^\pm(x)=\Gamma_5\varphi_0^\pm(x)
=\pm\varphi_0^\pm(x).
\end{equation}
We denote the number of $\varphi_0^+(x)$ ($\varphi_0^-(x)$) as 
$n_+$ ($n_-$).

\noindent
(ii) $\lambda_n\neq0$ and $\lambda_nf(\lambda_n^2)\neq\pm1$. As 
shown above,
\begin{equation}
H\varphi_n(x)=\lambda_n\varphi_n(x),\quad
H\widetilde\varphi_n(x)=-\lambda_n\widetilde\varphi_n(x),
\end{equation}
where
\begin{equation}
\widetilde\varphi_n(x)={1\over\sqrt{1-\lambda_n^2
f^2(\lambda_n^2)}}
\Gamma_5\varphi_n(x).
\end{equation}
We have
\begin{equation}
\Gamma_5\varphi_n(x)=\sqrt{1-\lambda_n^2f^2(\lambda_n^2)}
\widetilde\varphi_n(x),\quad
\Gamma_5\widetilde\varphi_n(x)=\sqrt{1-\lambda_n^2
f^2(\lambda_n^2)}
\varphi_n(x),
\end{equation}
and
\begin{eqnarray}
&&\gamma_5\varphi_n(x)=\sqrt{1-\lambda_n^2f^2(\lambda_n^2)}
\widetilde\varphi_n(x)+\lambda_nf(\lambda_n^2)\varphi_n(x),
\nonumber\\
&&\gamma_5\widetilde\varphi_n(x)=\sqrt{1-\lambda_n^2
f^2(\lambda_n^2)}
\varphi_n(x)-\lambda_nf(\lambda_n^2)\widetilde\varphi_n(x).
\end{eqnarray}

\noindent
(iii) $\lambda_nf(\lambda_n^2)=\pm1$, or
\begin{equation}
H\Psi_\pm(x)=\pm\Lambda\Psi_\pm(x),\quad\Lambda f(\Lambda^2)=1.
\end{equation}
We see
\begin{equation}
\Gamma_5\Psi_\pm(x)=0,
\end{equation}
and
\begin{equation}
\gamma_5\Psi_\pm(x)=\pm\Lambda f(\Lambda^2)\Psi_\pm(x)
=\pm\Psi_\pm(x).
\end{equation}
We denote the number of $\Psi_+(x)$ ($\Psi_-(x)$) as 
$N_+$ ($N_-$). From the relation $Tr\gamma_{5}=0$ valid on the 
lattice, one can derive the chirality sum rule\cite{chiu} 
\begin{equation}
n_{+}-n_{-}+N_{+}-N_{-}=0.
\end{equation}

We next recall the charge conjugation properties of various
operators.  
We employ the convention of the charge conjugation matrix $C$
\begin{eqnarray}
&&C\gamma^{\mu}C^{-1}=-(\gamma^{\mu})^{T},\\ 
&&C\gamma_{5}C^{-1}=\gamma^{T}_{5},\\ 
&&C^{\dagger}C=1,\ \  C^{T}=-C.
\end{eqnarray}
We then have\cite{fuji-ishi}\cite{suzuki}
\begin{eqnarray}
&&CDC^{-1}=D^{T},\ \ \ 
C\gamma_{5}\Gamma_{5}C^{-1}=(\gamma_{5}\Gamma_{5})^{T},
\nonumber\\
&&C\gamma_{5}H\gamma_{5}C^{-1}=H^{T},\ \ \ 
CH^{2}C^{-1}=(H^{2})^{T},
\nonumber\\
&&C(\gamma_{5}\Gamma_{5}\gamma_{5}
/\Gamma)C^{-1}=(\Gamma_{5}/\Gamma)^{T}
\end{eqnarray}
where 
\begin{equation}
\Gamma=\sqrt{\Gamma^{2}_{5}}=\sqrt{(\gamma_{5}\Gamma_{5}
\gamma_{5})^{2}}=\sqrt{1-H^{2}f^{2}(H^{2})}.
\end{equation} 
Here we imposed the relation  
$CDC^{-1}=D^{T}$ or $(CD)^{T}=-CD$ which is consistent 
with the defining algebraic relation (2): To be precise, we need
 to perform simultaneously a suitable charge conjugation of 
gauge field  in gauge theory\footnote{If one defines the CP
operation by $W=C\gamma_{0}=\gamma_{2}$ with hermitian 
$\gamma_{2}$ and the CP transformed gauge 
field by $U^{CP}$, one has $WD(U^{CP})W^{-1}=D(U)^{T}$. If the 
parity is realized in the standard way, one has 
$CD(U^{C})C^{-1}=D(U)^{T}$.}. In 
the following this charge 
conjugation of gauge field is implicitly assumed when we deal 
with theories with  gauge field.

We now examine the  CP symmetry in   chiral gauge theory 
\begin{equation}
{\cal L}=\bar{\psi}_{L}D\psi_{L}
\end{equation}
where we defined the projection operators
\begin{eqnarray}
&&D=\bar{P}_{L}DP_{L}+\bar{P}_{R}DP_{R},\nonumber\\
&&\psi_{L,R}=P_{L,R}\psi,\ \ \ \bar{\psi}_{L,R}=\bar{\psi}
\bar{P}_{L,R}.
\end{eqnarray}
If the algebra (2) is applicable to strong interactions it is 
natural that the parity operation is realized in the standard 
way, and we concentrate
on the charge conjugation. The proper transformation property 
under CP is then ensured by 
\begin{eqnarray}
&&CP_{L}C^{-1}=\bar{P}^{T}_{R}, \ \ \ C\bar{P}_{L}C^{-1}
=P^{T}_{R}
\end{eqnarray}
which transforms ${\cal L}\rightarrow {\cal L}^{c}
=\bar{\psi}_{R}D\psi_{R}$ under the charge conjugation 
$\bar{\psi}\rightarrow\psi^{T}C$ and 
$\psi\rightarrow-C^{-1}\bar{\psi}^{T}$.
One can confirm that the following projection operators
\begin{eqnarray}
P_{L,R}=\frac{1}{2}(1\mp \Gamma_{5}/\Gamma),\nonumber\\
\bar{P}_{L,R}=\frac{1}{2}(1\pm \gamma_{5}\Gamma_{5}\gamma_{5}
/\Gamma).
\end{eqnarray}
satisfy the requirement (32). 
One can in fact prove that these projection operators give the
unique solutions by using the  following statement.\\ 
{\bf Statement}:
If the operators $\widetilde U$ and $V$ are regular in 
$\gamma_5$ and $H$ and satisfy
\begin{equation}
\widetilde UH+HV=0,
\label{eq:one}
\end{equation}
and (with $B=C\gamma_5$, $B^T=-B$)
\begin{equation}
B\widetilde UB^{-1}=V^T,\quad BVB^{-1}=\widetilde U^T,
\label{eq:two}
\end{equation}
for a {\it generic} $H$, then they are of the form
\begin{equation}
\widetilde U=V=\Gamma_5h(H^2),
\label{eq:three}
\end{equation}
with a regular function $h(H^2)$.
\\
{\bf Proof}:
Since $\gamma_5=\Gamma_5+Hf(H^2)$, $\widetilde U$ and $V$ are 
regular also in $\Gamma_5$ and $H$. Noting 
$\Gamma_5^2=1-H^2f^2(H^2)$, the most general form
of $\widetilde U$ reads
\begin{equation}
\widetilde U=g(H)+\Gamma_5h(H^2)+\Gamma_5Hk(H^2).
\label{eq:seventeen}
\end{equation}
This implies
\begin{equation}
B\widetilde UB^{-1}=[g(H)+\Gamma_5h(H^2)-\Gamma_5Hk(H^2)]^T,
\label{eq:five}
\end{equation}
from $BHB^{-1}=H^T$, $B\Gamma_5B^{-1}=\Gamma_5^T$ and 
$\Gamma_5H+H\Gamma_5=0$.
Eqs.~(\ref{eq:two}) and (\ref{eq:five}) imply
\begin{equation}
V=g(H)+\Gamma_5h(H^2)-\Gamma_5Hk(H^2),
\end{equation}
and thus eq.~(\ref{eq:one}) imposes
\begin{equation}
Hg(H)+\Gamma_5H^2k(H^2)=0.
\end{equation}
The matrix element of this equation between 
$\widetilde\varphi_n(x)$
and $\varphi_n(x)$ reads
\begin{equation}
(\widetilde\varphi_n,[Hg(H)+\Gamma_5H^2k(H^2)]\varphi_n)
=\sqrt{1-\lambda_n^2f^2(\lambda_n^2)}\lambda_n^2k(\lambda_n^2)=0.
\end{equation}
This shows that $k(x)$ must have zero at $x=\lambda_n^2$, 
but this is impossible for a generic $H$ unless $k(x)=0$. 
Similarly, we 
have $g(H)=0$ and obtain eq.~(\ref{eq:three}).
\\

On the basis of this statement, one can construct the 
modified chiral operators
\begin{equation}
\Gamma_{5}/\Gamma, \ \ \ \gamma_{5}\Gamma_{5}\gamma_{5}/\Gamma
\end{equation}
with $(\Gamma_{5}/\Gamma)^{2}=1$ and 
$(\gamma_{5}\Gamma_{5}\gamma_{5}/\Gamma)^{2}=1$. 
In this construction, we assumed that $h(H^2)$ exhibits the 
most favorable property, namely, has no zeroes. 
These projection operators (33) however  inevitably contain  
singularities in the modified 
chiral operators $\Gamma_{5}/\Gamma$ and
 $\gamma_{5}\Gamma_{5}\gamma_{5}/\Gamma$, as is specified by 
our theorem. These projection operators (33) also become 
singular in the presence of topologically non-trivial 
gauge fields, since the massive modes $N_{\pm}$ in (22)
inevitably appear as is indicated  by the chirality sum rule 
(24).
This generalizes the analysis of Hasenfratz\cite{hasenfratz} in 
a more abstract setting.
\\

As for the analysis of the Majorana fermion, we start with
\begin{eqnarray}
{\cal L}
&=&\bar{\psi}_{R}D\psi_{R}+\bar{\psi}_{L}D\psi_{L} 
+ m[\bar{\psi}_{R}\psi_{L}+\bar{\psi}_{L}\psi_{R}]\nonumber\\
&&+ 2g[\bar{\psi}_{L}\phi\psi_{R}
+\bar{\psi}_{R}\phi^{\dagger}\psi_{L}]\\
&=&\bar{\psi}D\psi 
+ m\bar{\psi}\gamma_{5}\Gamma_{5}\psi
+ \frac{g}{\sqrt{2}}\bar{\psi}[A+
(\gamma_{5}\Gamma_{5}\gamma_{5}/\Gamma)A
(\Gamma_{5}/\Gamma)
+i(\gamma_{5}\Gamma_{5}\gamma_{5}/\Gamma)B
+iB(\Gamma_{5}/\Gamma)]\psi\nonumber
\end{eqnarray}
where we used $\phi=(A + iB)/\sqrt{2}$. 
We then  make the substitution
\cite{suzuki}\cite{nicolai}\cite{van nieuwenhuizen} 
\begin{eqnarray}
&&\psi=(\chi+i\eta)/\sqrt{2},\nonumber\\
&&\bar{\psi}=(\chi^{T}C-i\eta^{T}C)/\sqrt{2}
\end{eqnarray}
and obtain\footnote{If $(CO)^{T}=-CO$ 
for a general operator $O$, the cross
term vanishes $\eta^{T}CO\chi-\chi^{T}CO\eta=0$ by using the 
anti-commuting property of $\chi$ and $\eta$. In the presence of 
background gauge field, we assume that the representation of 
gauge symmetry is real. }
\begin{eqnarray}
{\cal L}
&=&\frac{1}{2}\chi^{T}CD\chi 
+ \frac{1}{2}m\chi^{T}C\gamma_{5}\Gamma_{5}\chi\nonumber\\
&&+ \frac{g}{2\sqrt{2}}\chi^{T}C[A+
(\gamma_{5}\Gamma_{5}\gamma_{5}/\Gamma)A
(\Gamma_{5}/\Gamma)
+i(\gamma_{5}\Gamma_{5}\gamma_{5}/\Gamma)B
+iB(\Gamma_{5}/\Gamma)]\chi\nonumber\\
&+&\frac{1}{2}\eta^{T}CD\eta
+ \frac{1}{2}m\eta^{T}C\gamma_{5}\Gamma_{5}\eta\nonumber\\
&&+ \frac{g}{2\sqrt{2}}\eta^{T}C[A+
(\gamma_{5}\Gamma_{5}\gamma_{5}/\Gamma)A
(\Gamma_{5}/\Gamma)
+i(\gamma_{5}\Gamma_{5}\gamma_{5}/\Gamma)B
+iB(\Gamma_{5}/\Gamma)]\eta.
\end{eqnarray} 
One can then define the Majorana fermion $\chi$ (or $\eta$) and 
the resulting Pfaffian.
But this formulation  of the Majorana fermion\cite{fuji-ishi}
inevitably suffers from the singularities of the modified 
chiral operators $\Gamma_{5}/\Gamma$ and 
$\gamma_{5}\Gamma_{5}\gamma_{5}/\Gamma$ in the Brillouin zone 
without gauge fields or from the singularities of these 
chiral operators caused by the massive modes $N_{\pm}$ (22) in 
the presence of topologically non-trivial gauge fields.

We note that the condition (32), which is required by the 
consistent $CP$ property in (30), is directly related to the 
condition of the consistent Majorana reduction for the term 
containing scalar field  $A(x)$,
\begin{equation} 
C(\gamma_{5}\Gamma_{5}\gamma_{5}/\Gamma)A(x)
(\Gamma_{5}/\Gamma)=-[C(\gamma_{5}\Gamma_{5}\gamma_{5}/\Gamma)
A(x)(\Gamma_{5}/\Gamma)]^{T}
\end{equation}
in the Yukawa coupling, if one recalls that the difference 
operators in $\Gamma_{5}$ and $\Gamma$ do not commute with 
the field $A(x)$. In other words, if one uses the projection 
operators which do not satisfy the condition (32)
the consistent Majorana reduction is not realized. 
For the chiral symmetric Yukawa
couplings such as in supersymmetry  the Majorana reduction is 
thus directly related 
to the condition (32) (and consequently to the CP invariance  
of (43)), provided that parity properties are the standard 
ones\footnote{ The Yukawa coupling in the Higgs mechanism on the 
lattice in general is also constrained by CP invariance. In this 
sense, the constraint arising from CP symmetry is more 
universal.}. 
 
It should be noted that our analysis does not
show how serious the complications associated with CP
symmetry and Majorana reduction are in the actual applications 
of lattice regularization. As for the breaking of CP symmetry, 
it could be similar to the breaking of Lorentz symmetry for
finite lattice spacing $a$; it may well be restored in a 
suitable continuum limit. Nevertheless it must be useful to keep 
in mind that exact and manifest CP is not implemented for a 
general Ginsparg-Wilson operator. 
As for the absence of Majorana 
reduction, a reliable analysis of supersymmetry in some of 
supersymmetric theories is not possible; 
practically this could be serious since the divergence 
cancellation in supersymmetric theory, for example, is very 
sensitive to the precise implementation of supersymmetry algebra.

In conclusion, we have shown that both of 
the consistent definitions of CP symmetry in chiral gauge 
theory\cite{hasenfratz} and the Majorana fermion in the 
presence of chiral symmetric Yukawa couplings\cite{fuji-ishi} 
are based on the same condition (32), and that  
the construction of projection operators (33) inevitably 
suffers from the singularities in the modified chiral 
operators for any Dirac operator $D$ satisfying the algebraic 
relation (2). 
Our analysis is based on several key  
assumptions including the algebraic relation (2) itself, 
though those assumptions appear to 
be natural in the framework of Ginsparg and Wilson.  
We find it quite interesting that the breaking of CP
symmetry and a conflict with Majorana reduction are directly 
related to the basic notions of locality and species doubling
in lattice theory.
A detailed analysis of the possible implications of the breaking 
of CP symmetry and the absence of Majorana reduction will be 
given elsewhere.

\end{document}